\documentclass[prd,superscriptaddress,amsfonts,amssymb,amsmath,showpacs,,twocolumn]{revtex4-2}
\usepackage{bm}
\usepackage{amsfonts}
\usepackage{latexsym}
\usepackage[latin1]{inputenc}
\usepackage{graphicx}
\usepackage{amsmath}
\usepackage{palatino}
\usepackage{mathpazo}
\usepackage{textcomp} 
\usepackage{float}
\usepackage{booktabs}
\usepackage{dcolumn}
\usepackage{ragged2e}
\usepackage{hyperref}
\hypersetup{colorlinks,citecolor=green}
\hypersetup{colorlinks=true,linkcolor=blue,filecolor=blue,    urlcolor=magenta}
\usepackage{amsmath}
\usepackage{xcolor}
\usepackage{orcidlink}
\usepackage[caption=false]{subfig}
\usepackage{commath}
\usepackage[autostyle]{csquotes}
\def\jnl@style{\it}
\def\aaref@jnl#1{{\jnl@style#1}}

\def\aaref@jnl#1{{\jnl@style#1}}

\def\aj{\aaref@jnl{AJ}}                   
\def\apj{\aaref@jnl{ApJ}}                 
\def\apjl{\aaref@jnl{ApJ}}                
\def\apjs{\aaref@jnl{ApJS}}               
\def\apss{\aaref@jnl{Ap\&SS}}             
\def\aap{\aaref@jnl{A\&A}}                
\def\aapr{\aaref@jnl{A\&A~Rev.}}          
\def\aaps{\aaref@jnl{A\&AS}}              
\def\mnras{\aaref@jnl{Mon.~Not.~Roy.~Astron.~Soc.}}             
\def\prd{\aaref@jnl{Phys.~Rev.~D}}        
\def\plb{\aaref@jnl{Phys.~Lett.~B}}        
\def\prc{\aaref@jnl{Phys.~Rev.~C}}  
\def\prl{\aaref@jnl{Phys.~Rev.~Lett.}}    
\def\qjras{\aaref@jnl{QJRAS}}             
\def\skytel{\aaref@jnl{S\&T}}             
\def\ssr{\aaref@jnl{Space~Sci.~Rev.}}     
\def\zap{\aaref@jnl{ZAp}}                 
\def\nat{\aaref@jnl{Nature}}              
\def\aplett{\aaref@jnl{Astrophys.~Lett.}} 
\def\apspr{\aaref@jnl{Astrophys.~Space~Phys.~Res.}} 
\def\physrep{\aaref@jnl{Phys.~Rep.}}      
\def\physscr{\aaref@jnl{Phys.~Scr}}       
\def\commat{\aaref@jnl{Comm.~Math.~Phys.}}              
\def\science{\aaref@jnl{Science}}               
\def\cqg{\aaref@jnl{Classical Quant.~Grav.}}            
\def\jpcs{\aaref@jnl{JPCS}}                                     
\def\ijmpd{\aaref@jnl{Int.~J.~Mod.~Phys.~D}}                    
\def\grg{\aaref@jnl{Gen.~Relat.~Gravit.}}               
\def\rpp{\aaref@jnl{Rep.~Prog.~Phys.}}          
\def\npa{\aaref@jnl{Nucl.~Phys.~A}}        
\def\lrr{\aaref@jnl{Living Rev.~Rel.}}                   
\def\jcap{\aaref@jnl{J.~Cosmology Astropart.~Phys.}}    
\def\rmp{\aaref@jnl{Rev.~Mod.~Phys.}}   
\def\epjc{\aaref@jnl{Eur.~Phys.~J.~C}}

\allowdisplaybreaks[1]

\begin{document}

\color{black}       

\title{Properties of white dwarf with anisotropic pressure
in Rainbow gravity }

\author{Takol Tangphati} 
\email{takoltang@gmail.com}
\affiliation{School of Science, Walailak University, Thasala, \\ Nakhon Si Thammarat, 80160, Thailand}

\author{Grigoris Panotopoulos}
\email{grigorios.panotopoulos@ufrontera.cl}
\affiliation{Departamento de Ciencias F{\'i}sicas, Universidad de la Frontera, Casilla 54-D, 4811186 Temuco, Chile}

\author{Ayan Banerjee} 
\email{ayanbanerjeemath@gmail.com}
\affiliation{Astrophysics and Cosmology Research Unit, School of Mathematics, Statistics and Computer Science, University of KwaZulu--Natal, Private Bag X54001, Durban 4000, South Africa}

\author{Anirudh Pradhan} 
\email[]{pradhan.anirudh@gmail.com}
\affiliation{Centre for Cosmology, Astrophysics and Space Science, GLA University, Mathura-281 406, Uttar Pradesh, India}


\date{\today}

\begin{abstract}
We investigate the properties of anisotropic white dwarf stars within the rainbow gravity adopting for matter content the Chandrasekhar model based on an ideal Fermi gas at zero temperature.  We study in detail the effects of the anisotropic factor on stellar mass and radius, the speed of sound, and the relativistic adiabatic index in both radial and tangential directions. We find that causality is never violated, whereas the stability criterion based on the relativistic adiabatic index is not met when the objects are characterized by a positive anisotropic factor close to the Chandrasekhar limit. We present this significant observation here for the first time, to the best of our knowledge.
\end{abstract}

\maketitle

\section{Introduction}

Einstein's theory of gravity has achieved remarkable success for over a century. The GR stands out as one of the best-tested 
theories on relatively small astronomical scales, such as the Solar System and compact astrophysical objects. Despite all of GR's success, the discovery of the accelerated expansion of the universe led to a renaissance in scientific society. As we know, gravity is an attractive force that causes the universe and all of its matter to contract. So the expansion of the universe would gradually slow down at a rate determined by the density of matter and energy within it. To overcome this limitation, it is necessary to assume additional exotic matter components (known as Dark Energy) within the framework of GR. However, the true nature of dark energy remains obscure, and an understanding of its physical properties is incomplete. 

Beside that, another possible explanation involves modifying the Einstein-Hilbert action, commonly known as modified gravitational theory, without requiring the introduction of dark components. Such modifications to GR include the adding of extra dimensions like Brane-World gravity \cite{Maartens:2010ar}, 
non-minimal geometry-matter coupling theories $f (R, \mathcal{L}_m)$  \cite{Harko:2010mv} and $f(R, T)$ \cite{Harko:2011kv}, higher-curvature theories e.g., $f(R)$  gravity \cite{Felice2010} and Lanczos-Lovelock gravity \cite{Padmanabhan:2013xyr}. Furthermore, massive gravity~\cite{Rham2011}, Brans-Dicke gravity~\cite{Brans1961} and  $f(Q)$ gravity \cite{BeltranJimenez:2017tkd} are also well-defined modified gravity theories from the theoretical as well as experimental/observational point of view.

Following this direction, in Ref. \cite{Magueijo:2002xx}, a generalization from the doubly special relativity was proposed, called `rainbow gravity'. This means that
particles with different energies distort the spacetime differently, and the usual Einstein's equations is modified as 
a one parameter family of equations. This alternative approach has received a lot of attention among researchers,  and we 
shall review the definition of rainbow metrics in the next section. Researchers have extensively studied the 
properties of static spherically symmetric black hole solutions within the framework of rainbow gravity
\cite{Ling:2005bp,Liu:2007fk,Garattini:2009nq,Gim:2014ira,Feng:2017gms,Dehghani:2018svw}.  In this context, the study of gravastar solutions
 have been found in \cite{Debnath:2019eor,Barzegar:2023ueo}. The authors of \cite{Garattini:2012ki} investigated the wormhole solution for a static and spherically symmetric spacetime. Additionally, Tudeshki \textit{et al} \cite{Tudeshki:2022wed} have studied an anisotropic dark energy and investigated their dynamical stability. Subsequently, dynamically stable neutron stars  and quark stars were demonstrated (for a detailed discussion, see, e.g., Refs. \cite{Hendi:2015vta,Tangphati:2023fey,Tangphati:2024qkd,Banerjee:2024inf}). Rainbow gravity holds significant importance as the spacetime depends on the form of the rainbow functions and thereby the energy of the probe particle that affects the spacetime background.

 Based on the above discussion, this paper develops a framework for discussing the white dwarfs in rainbow gravity, which are among the most extensively studied stars in stellar astrophysics. White dwarf stars, in astronomy, 
are the endpoints of the evolution of a low or medium mass stars. Therefore, scientists believe that white dwarfs are the most informative objects on the stellar evolution theory to
measure the age of stellar populations, the kinematics, and the star formation history of our galaxy. Interestingly, the existence
of white dwarfs cannot be more massive than 1.4 $ M_{\odot}$. This limit is called the Chandrasekhar limit for the maximum mass of a stable white dwarf star \cite{Chandrasekhar1931,Chandrasekhar1935}. The theoretical evolution of white dwarfs is well understood in \cite{Corsico:2012wd,Camisassa:2022pet}.  However,  recent detection of some extremely luminous type-Ia supernovae, such as SN 2007if, SN 2006gz, SN 2003fg and SN 2009DC~\cite{Howell2006,Scalzo2010} has cast the question on the
maximum mass of white dwarfs. This  being known as ``super-Chandrasekhar" white dwarfs \cite{Hicke2007,Yamanaka2009,Silverman2011,Taubeberger2011}. Researchers have extensively studied this class of objects in a variety of settings, such as super strong uniform magnetized white dwarfs~\cite{Das2012,Das2013,Das2014,Franzon2015,Deb2022}, electrical charged white dwarfs~\cite{Liu2014,Carvalho2018} and rotating white dwarfs~\cite{Boshkayev2011,Boshkayev2012}.

Meanwhile, local anisotropy can exist within a self-gravitating systems \cite{Herrera:1997plx}. In Refs. \cite{Ruderman:1972plx,Canuto:1974plx} 
authors have show that the nuclear matter may be anisotropic at
very high densities where the nuclear interactions must be treated relativistically. Some theoretical investigations have suggested that the existence of anisotropic pressure can be triggered by various phenomena, e.g., relativistic nuclear interactions \cite{Ruderman:1972plx,Canuto:1974plx}, superfluid cores \cite{Heiselberg:1999mq},
strong magnetic fields \cite{Cardall:2000bs,Ioka:2003nh,Ciolfi:2010td}, etc. Considering pressure anisotropy,  one can construct relativistic stars \cite{Herrera:2015vca,Maurya:2017uyk} 
or such exotic objects as wormholes \cite{Morris:1988cz,Visser:2003yf}. Surprisingly, white dwarfs with an anisotropy matter source in GR is poorly studied. Our purpose will be to demonstrate the effects of modified gravity and fluid anisotropy on the internal structure of white dwarfs.

In addition, white dwarfs in modified theories of gravity have also attracted much attention among researchers, see Refs. \cite{Das2015,Jing2016,Banerjee2017,Carvalho17,Kalita2018,Panah2019,Liu2019,Rocha2020,Wojnar2021,Kalita2021,Kalita2022}. Recently, white dwarfs within the framework of Rastall-Rainbow gravity have been evaluated in \cite{Li:2023fux}. In dRGT massive gravity, Panah and Liu \cite{EslamPanah:2018evk} have studied white dwarfs, and showed that the maximum
 mass can be more than the Chandrasekhar limit ($M >1.45 M_{\odot}$). From the foregoing outline, in this paper we are going to investigate the possible existence and analyze the stability of white dwarfs in Rainbow gravity. With the same modified gravity, our interest lies in emphasizing the effect of Rainbow function on the structural properties of white dwarfs.  In addition, we compute the static stability criterion, the adiabatic index, and the sound velocity. Finally, we compare our results with GR counterpart.

 The paper is organized as follows:  Section \ref{sec2} provides a brief review of the Rainbow gravity and derives the modified TOV equations for stellar structure. In Section \ref{sec2}, we set up the equations
 of state for white dwarfs including the structure equations for anisotropic fluid distribution. In Section \ref{sec4}, we solve the modified TOV equations numerically and demonstrate the effects of model parameters on the mass-radius relations. Section \ref{sec5} is devoted to computing the dynamic stability of white dwarfs in this gravity theory.  Concluding remarks are given in Section \ref{sec6}.

\section{Review of Gravity's Rainbow and stellar structure equations}\label{sec2}

\subsection{Rainbow theory}
\label{TOV_RR}

Among the many challenges, the biggest challenge in theoretical physics is how to unify quantum mechanics and general relativity (GR) together. To address this challenge, physicists have proposed a number of approaches, such as loop quantum gravity, string theory, non-commutative geometry, and so on. Among them deformed (or doubly) special relativity (DSR) has been conceived as an extension of special relativity \cite{Amelino-Camelia:2000stu}
(also refer to \cite{Kowalski-Glikman:2006ssl}). This modified dispersion relation assume the existence of two observer-independent scales: the speed of light and the Planck energy. However,  DSR already faced the most pressing problem is the 'soccer ball' problem.
In an attempt to provide possible answers, Magueijo and Smolin \cite{Magueijo:2002xx} proposed a new theory called `rainbow gravity' (RG) as a  generalization of DSR to include curvature in spacetime. Within this formalism the geometry of spacetime depends on the energy of the test particle and leads to the energy-dependent distortions in spacetime. Specifically, given a modified dispersion relation for a particle of mass $m$, which is represented by the following equation: 
\begin{equation}
E^{2} \Xi(x)^{2} - p^{2}\Sigma(x)^{2} = m^{2},
\label{eq1}
\end{equation}
where $\Xi(x)$ and $\Sigma(x)$ are functions of dimensionless ratio $x = E/E_p$ with $E$ is the particle's total energy and $E_p$, the Planck energy, is defined as $E_p = \sqrt{\frac{\hslash c^5}{G}}$. These functions are known as rainbow functions and have an influence the ultraviolet regime \cite{Dehghani:2018svw}.  In the limit of $x \rightarrow 0$, the rainbow functions with the following properties: 
\begin{equation}
    \lim_{x\rightarrow 0} \Xi(x)=1, \quad \lim_{x \rightarrow 0} \Sigma(x)=1,
    \label{eq2}
\end{equation}
thereby recovering the standard energy dispersion relation. Following \cite{Magueijo:2002xx}, the rainbow metric which is given
 in terms of a one-parameter family of orthonormal frame
 fields 
\begin{equation}
g^{\mu\nu}(x)=\eta^{ab} e_{a}^{\mu}(x)\otimes e_{b}^{\nu}(x),
\label{eq3}
\end{equation}
where the vierbeins $e_{a}^{\mu}(x)$ vary with energy:
\begin{equation}
e_{0}^{\mu}(x)=\frac{1}{\Xi(x)} \widetilde{e}_{0}^{\mu}, \quad e_{k}^{\mu}(x)=\frac{1}{\Sigma(x)} \widetilde{e}_{k}^{\mu},
\label{eq4}
\end{equation}
where the `tilde' quantities represent the energy-independent tetrads.  As a consequence,  the usual Einstein's equations
is modified as a one parameter family of equations,
\begin{equation}
G_{\mu\nu}(x) \equiv R_{\mu \nu}(x) - \frac{1}{2}g_{\mu \nu}(x)R(x) = k(x)T_{\mu \nu}(x),
\label{eq10}
\end{equation}
where $k(x) = 8 \pi G(x)$ with $G_{\mu \nu}(x)$ is the Einstein tensor and  $T_{\mu \nu}(x)$ is the stress-energy tensor, the source of spacetime curvature. Here, we use a system of units in which
 $G(x)=1$ for simplicity.

\subsection{Modified TOV equations of rainbow gravity}

Here, we demonstrate a solution for WDs to the modified field equations (\ref{eq10}), and thereby conventional spherically symmetric metric is replaced by a rainbow metric parameterized by the energy of the test particle \cite{Magueijo:2002xx},
\begin{equation} \label{metric}
    ds^{2} = -\frac{e^{2\Phi(r)}}{\Xi^{2}(x)} dt^{2} + \frac{e^{2\lambda(r)}}{\Sigma^{2}(x)} dr^2 + \frac{r^{2}}{\Sigma^{2}(x)}(d\theta^{2} + \sin^2{\theta}d\phi^{2}),
\end{equation}
 where the metric functions $\Phi(r)$ and $\lambda(r)$  are function of the radial coordinate $r$ only. In addition, the rainbow functions $\Xi(x)$ and $\Sigma(x)$ are invariant with respect to the spacetime coordinates ($r$, $t$, $\theta$, $\phi$). 
 
 Let us now consider that the matter content is described by an anisotropic fluid with energy-momentum tensor given
by  
\begin{equation}\label{eq12}
T_{\mu\nu} = (\rho + p_t)u_\mu u_\nu + p_t g_{\mu\nu} - (p_t - p_r) \chi_{\mu}\chi_{\nu},
\end{equation}
 where $\rho(r)$ is the energy density, $p_r(r)$ is the radial pressure, and $p_t(r)$ is the transverse pressure, respectively.
The fluid 4-velocity is written as
\begin{equation}
u^\mu = \left(\frac{\Xi(x)}{e^{\Phi(r)}}, 0, 0, 0\right),
\end{equation}
with the condition $u_\mu u^\mu = -1$.  The $\chi_{\mu}$ is the unit radial vector with $\chi_{\mu}\chi^{\mu}=1$.

The form of the spacetime metric \eqref{metric}, together with the energy-momentum tensor \eqref{eq12}, we obtain the modified Tolman-Oppenheimer-Volkoff (TOV) equations  \eqref{eq10} with the following forms \cite{Hendi:2015vta}:
\begin{eqnarray}
&&  M_{\rm{eff}}(r, x) =  \int^{r}_{0} \frac{4 \pi r^2 \rho(r)}{\Sigma^{2}(x)} dr \equiv \frac{m(r)}{\Sigma^{2}(x)},  \label{eq14} \\
&&  p_{r}' = -(\rho + p_{r})\Phi' + \frac{2}{r}(p_{t} - p_{r}),  \label{eq15} \\
&&  \Phi'(r) = \frac{M_{\rm{eff}}(r, x) \Sigma^{2}(x) + 4\pi r^3 p_r(r)}{r(r - 2M_{\rm{eff}}(r, x))\Sigma^{2}(x)},  \label{eq16}
\end{eqnarray}
where the prime denotes a derivative with respect to the radial coordinate, $r$. The three differential equations above contain five unknowns, so we should specify an equations of state (EoS) to provide a comprehensive description of the configuration under
consideration. We will discuss the topics below in detail.

\section{Equation of state} \label{sec3}

As a way to solve the modified TOV equations, it is necessary to provide an EoS that establishes a relationship  between the pressure and density of the system. In this context, we consider the Chandrasekhar EoS~\cite{Chandrasekhar1935}, 
which can be expressed in the following form
\begin{eqnarray}
&&p_r(k_{F})=\frac{1}{3\pi^{2}\hbar^{3}}\int^{k_{F}}_{0} \frac{k^{4}}{\sqrt{k^{2}+m_{e}^{2}}}dk \nonumber\\
&&~~~~~~~~~~~~~ =\frac{\pi m_{e}^{4}}{3 h^{3}}\Big[x_{F}(2x_{F}^{2}-3)\sqrt{x_{F}^{2}+1}+3\sinh^{-1}x_{F}\Big]~,\label{Eos-P}\\
&&~~~\rho=\frac{8\pi\mu_{e}m_{H}m_{e}^{3}}{3h^{3}}x_{F}^{3}~,\label{Eos-ro}
\end{eqnarray}
where $\hbar=h/2\pi$, $h$ is the Plank's constant, $x_{F} \equiv p_{F}/m_{e}c$, $p_{F}$ is the Fermi momentum, $k$ is the momentum of electrons, $\mu_{e}$ is the mean molecular weight per electron (we choose $\mu_{e} = $2 for our work), $m_{e}$ is the electron mass and $m_{H}$ is the mass of hydrogen atom, respectively. \\
\\ 
Besides the Chandrasekhar EoS, we also adopt Quasi-Local (QL) model for an anisotropic matter distribution proposed by Horvat \textit{et al} \cite{Horvat:2010xf}, which is described by the following EoS
\begin{eqnarray}
    \sigma &\equiv& p_{\perp} - p_r =  \beta p_r \mu,
    \label{anisotropy}
\end{eqnarray}
where $\beta$ is a constant that measures the degree of anisotropy in the fluid and $\mu  = 2m(r)/r$ represents the local compactness of the star. The parameter $\beta$ is restricted to the domain $[-2, 2]$, see Refs. \cite{Doneva:2012rd, Silva:2014fca, Yagi:2015hda, Pretel:2020xuo, Rahmansyah:2020gar, Rahmansyah:2021gzt, Folomeev:2018ioy,Tangphati:2022cqr} for more. We note here that the choice $r\to 0$, the effect of anisotropy vanishes at the stellar interior i.e., $\sigma =0$, and thereby recovering the isotropic solution. The choice of (\ref{anisotropy}) also allows for the vanishing of pressure components at the star's surface, i.e., $p_r \left( r \rightarrow R \right) = p_{\perp} \left( r \rightarrow R \right) = 0$.  


\begin{figure}
    \centering
    \includegraphics[width = 8 cm]{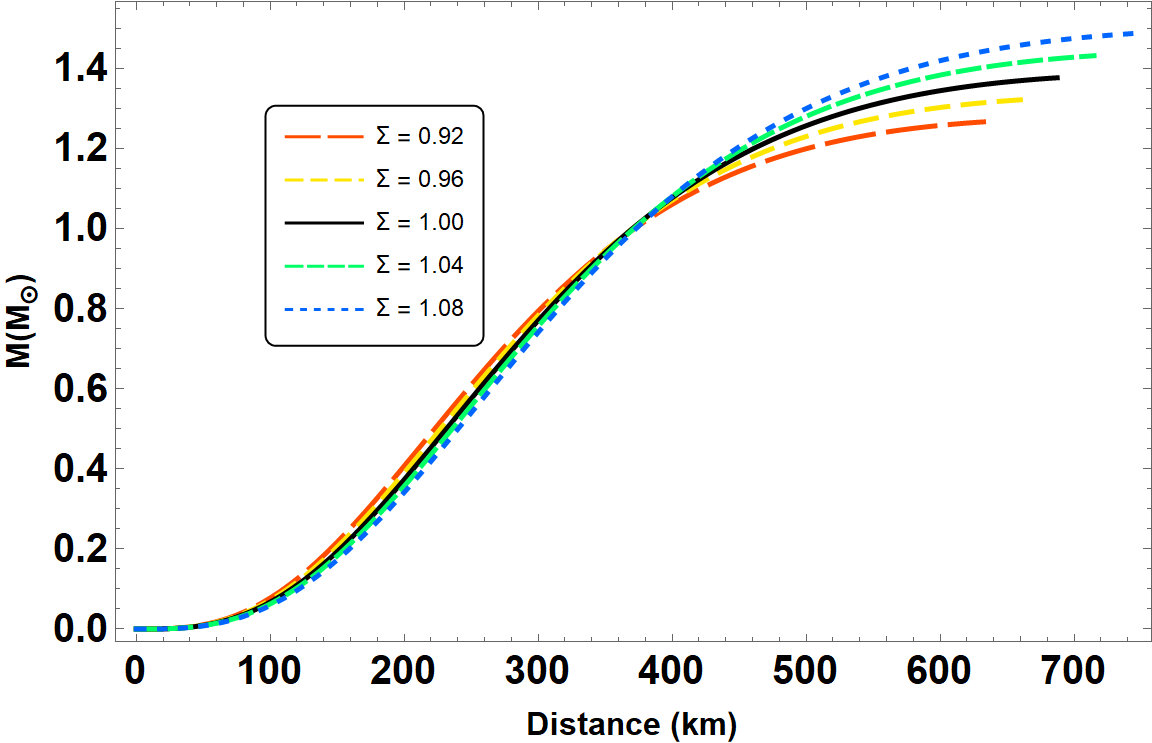}
    \includegraphics[width = 8 cm]{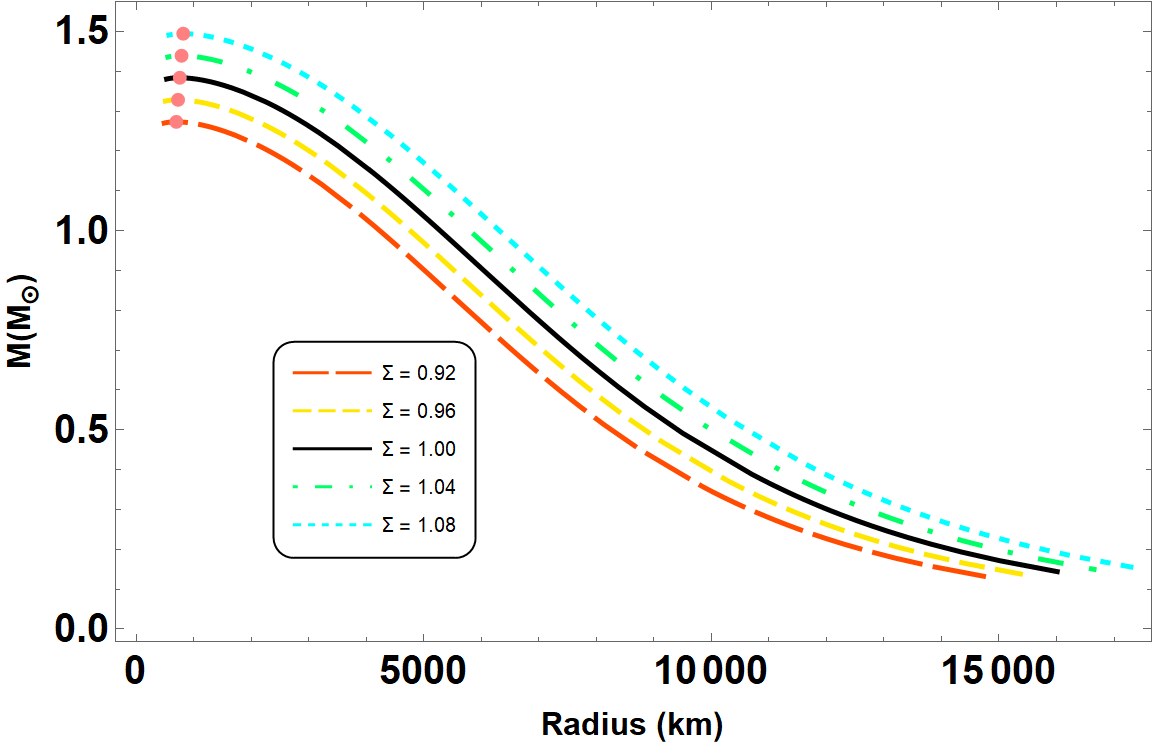}
    \includegraphics[width = 8 cm]{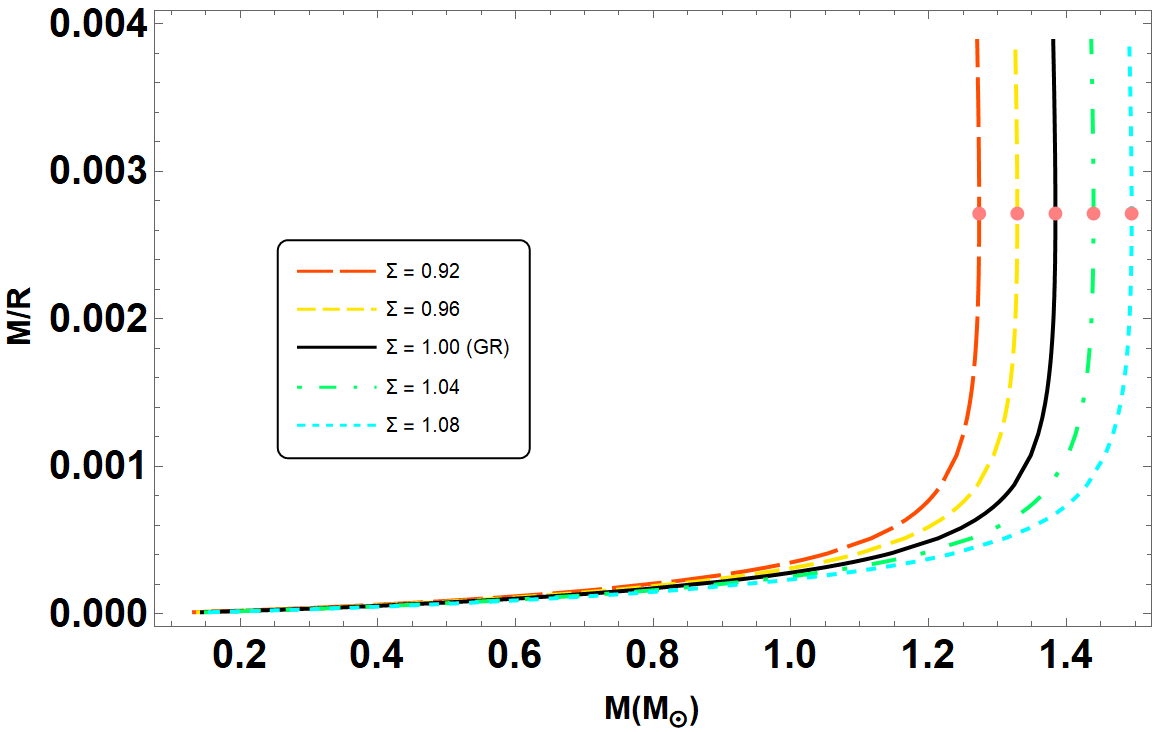}
    \caption{Mass function versus radial coordinate (upper panel), mass-to-radius relationship (panel in the middle) and factor of compactness versus stellar mass (lower panel) of white dwarf stars  within rainbow gravity, taking into account  variation of} $\Sigma \in [0.92, 1.08]$ for an anisotropic level $\beta = 0.1$. The black solid line illustrates the general relativity scenario for $\Sigma = 1$. The used model parameters are shown in Table \ref{tab_varyLambda}.
    \label{vary_Sigma}
\end{figure}


\begin{figure}
    \centering
    \includegraphics[width = 8 cm]{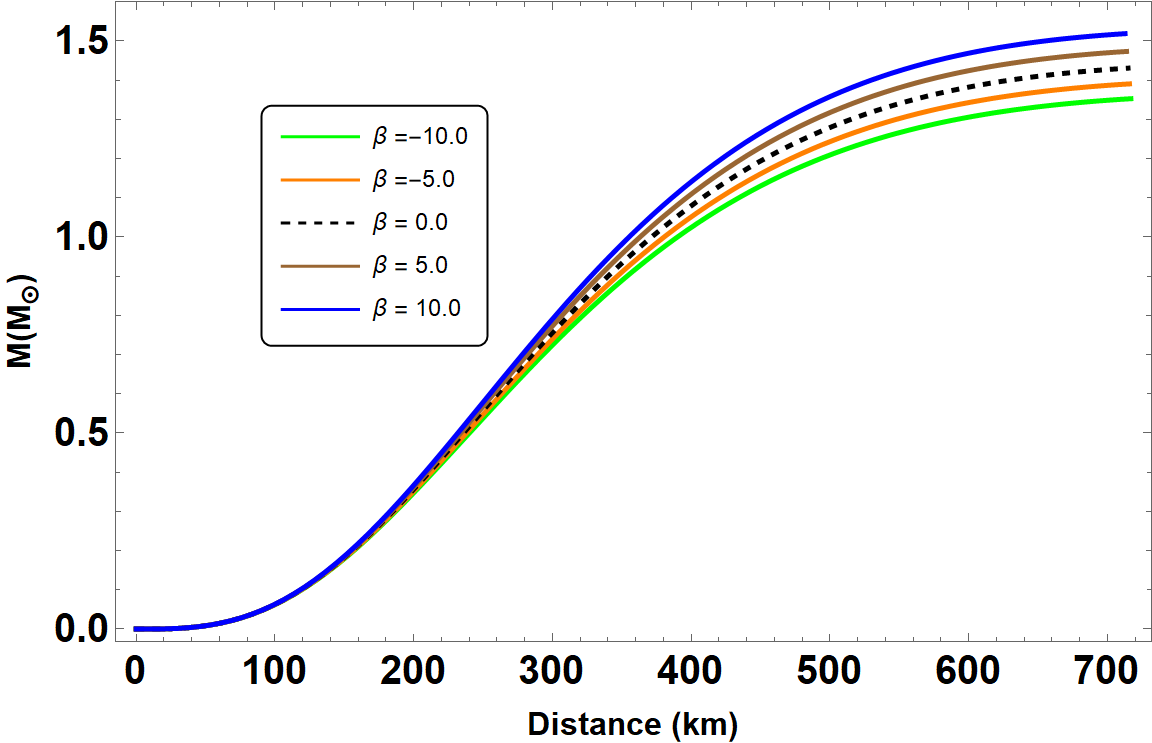}
    \includegraphics[width = 8 cm]{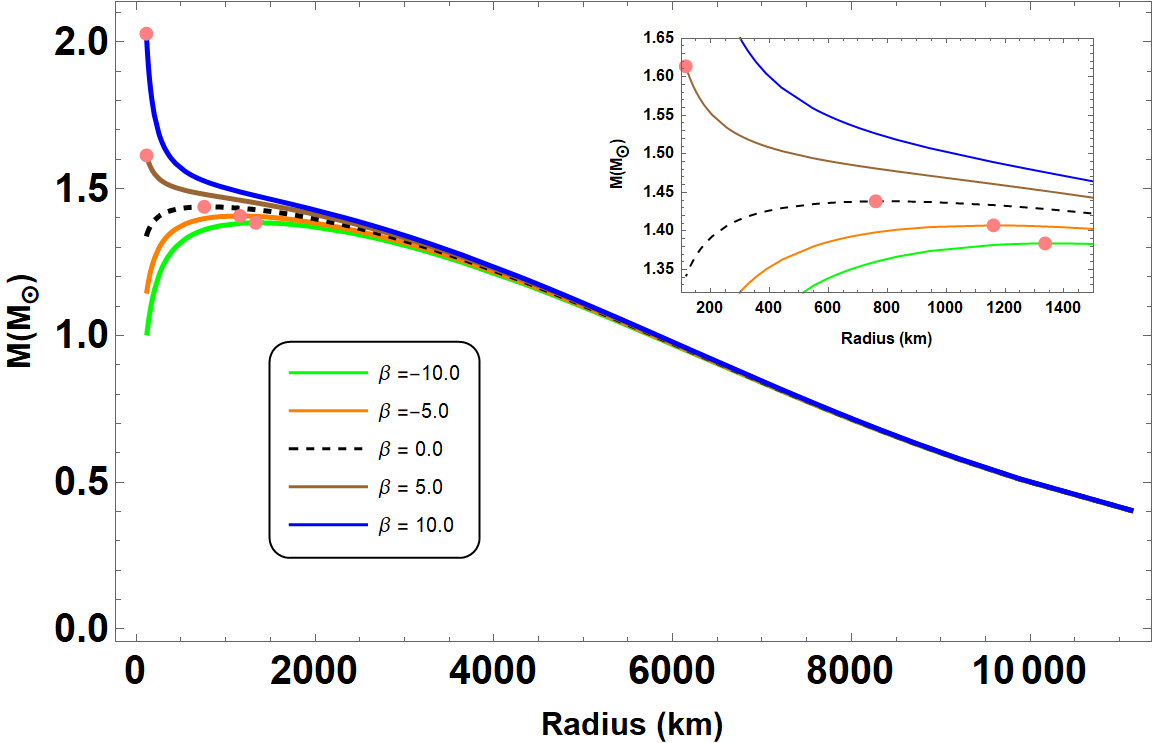}
     \includegraphics[width = 8 cm]{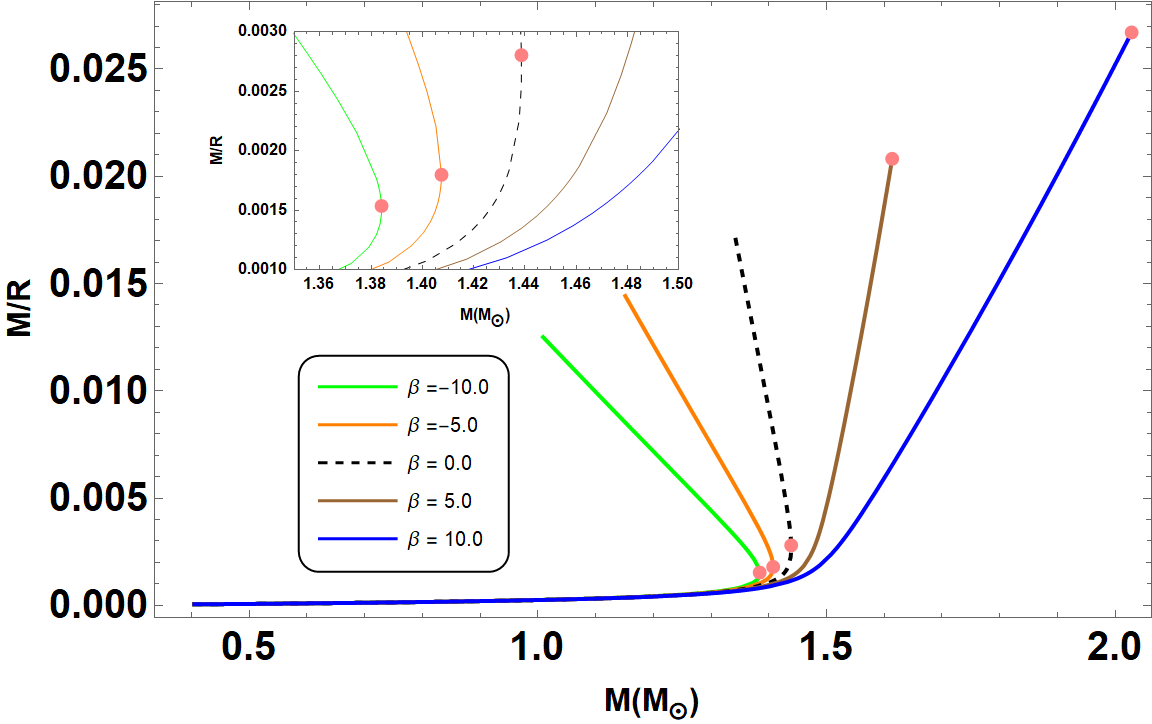}
    \caption{
     Mass function versus radial coordinate (upper panel), mass-to-radius relationship (panel in the middle) and factor of compactness versus stellar mass (lower panel) of white dwarf stars within rainbow gravity, taking into account variations of $\beta\in [-10, 10]$  for the rainbow function $ \Sigma = 0.5$. The black dashed line presents the case of isotropic solution ($\beta = 0$) in rainbow gravity. The used model parameters are shown in Table \ref{vary_beta}.
    }
    \label{vary_beta}
\end{figure}


\begin{table}[h]
    \caption{Main characteristics of white dwarf stars in gravity's rainbow. Used model parameters are $\Sigma \in [0.92, 1.08]$ with $\beta = 0.1$.}
    \begin{ruledtabular}
    \begin{tabular}{ccccc}
$\Sigma$ & $M_{max}$     & R     & $\rho_c$ & $M/R$\\
        & $(M_{\odot})$ & (km)  & $\times 10^{12}$ kg/m$^3$ & \\
\hline
  0.92  & 1.27 & 695 & 33.69 & 0.0027 \\[0.03cm]
  0.96  & 1.33 & 725 & 33.69 & 0.0027 \\[0.03cm]
  1.00  & 1.38 & 756 & 33.69 & 0.0027 \\[0.03cm]
  1.04  & 1.44 & 786 & 33.69 & 0.0027 \\[0.03cm]
  1.08  & 1.49 & 816 & 33.69 & 0.0027 \\[0.03cm]
\end{tabular}
    \end{ruledtabular}    
    \label{tab_varyLambda}
\end{table}


\begin{table}[h]
    \caption{Main characteristics of white dwarf stars in gravity's rainbow. Used model parameters are $\beta\in [-10, 10]$  with $ \Sigma = 0.5$.}
    \begin{ruledtabular}
    \begin{tabular}{ccccc}
$\beta$ & $M_{max}$     & R     & $\rho_c$ & $M/R$\\
        & $(M_{\odot})$ & (km)  & $\times 10^{12}$ kg/m$^3$ & \\
\hline
  -10 & 1.38 & 1,336 & 6.40 & 0.0015 \\[0.03cm]
  -5  & 1.41 & 1,160 & 10.0 & 0.0018 \\[0.03cm]
   0  & 1.44 & 761   & 37.0 & 0.0028 \\[0.03cm]
   5  & 1.61 & -     & -    & -    \\[0.03cm]
   10 & 2.02 & -     & -    & -    \\[0.03cm]
    \end{tabular}
    \end{ruledtabular}    
    \label{tab_varysigma}
\end{table}


\section{Computational Setup} \label{sec4}

In the present study, we solve numerically the structure equations provided by Eqs. (\ref{eq14}) to (\ref{eq16}), supplemented  by the Chandrasekhar EoS and quasi-local model for an anisotropic matter distribution.  We impose at the center of the star appropriate initial conditions as follows:
\begin{eqnarray}
\rho(0)=\rho_{c}~~\text{and}~~m(0)=0~,\label{initial}
\end{eqnarray}
where $\rho_{c}$ is the central energy density.  Those initial conditions ensure regularity of the interior solution describing hydrostatic equilibrium. We then proceed with the integration  throughout the star until we reach its surface, where we impose the matching conditions as follows:
\begin{equation}
p_r(R) = 0 = p_t(R), \; \; \; \; \;\; M = m(R).
\end{equation}
 Those matching conditions permit us to determine both the radius, $R$, and the mass, $M$, of the star.

The model discussed here is based on the variation of two free parameters, $\beta$ and $ \Sigma$. The first measures anisotropy strength, while the second measures deviation from GR. When $\beta=0$ the objects are isotropic, while when $\Sigma=1$ we recover Einstein's gravity.

In Fig. \ref{vary_Sigma} we show the impact of $\Sigma$ variations on stellar mass, radius and factor of compactness, $c=M/R$, assuming $\beta=0.1$, see Table \ref{tab_varyLambda} for more details. We obtain the mass function versus radial coordinate (upper panel), mass-to-radius relationship (panel in the middle) and factor of compactness versus stellar mass (lower panel). In all three panels, the solid black curve in the middle corresponds to the GR solution. The mass (in solar masses) decreases with the stellar radius (in km), which is in agreement with the solution of the Lane-Emden equation for non-relativistic objects when the polytropic index $n=3/2$. Moreover, the factor of compactness slightly increases with the stellar mass, although it remains small, around two orders of magnitude lower than that of neutron stars. $\Sigma < 1$ implies less massive stars, whereas $\Sigma > 1$ implies more massive objects.

The three panels of Fig. \ref{vary_beta} are similar to the ones of the previous figure, although this time we fix $\Sigma=0.5$ and we vary $\beta$, see Table \ref{tab_varysigma} for more details. A negative anisotropic factor implies less massive objects, whereas when the anisotropy is positive the stars become more massive.

\section{Stability Investigation of QS\lowercase{s}} \label{sec5}

Now we confront our results with the stability of white dwarf stars using the static stability criterion, the adiabatic index, and the sound velocity. The methods used for assessing the stability of these stars are briefly described below.

\subsection{Static Stability Criterion}

\begin{figure}
    \centering
 \includegraphics[width = 8 cm]{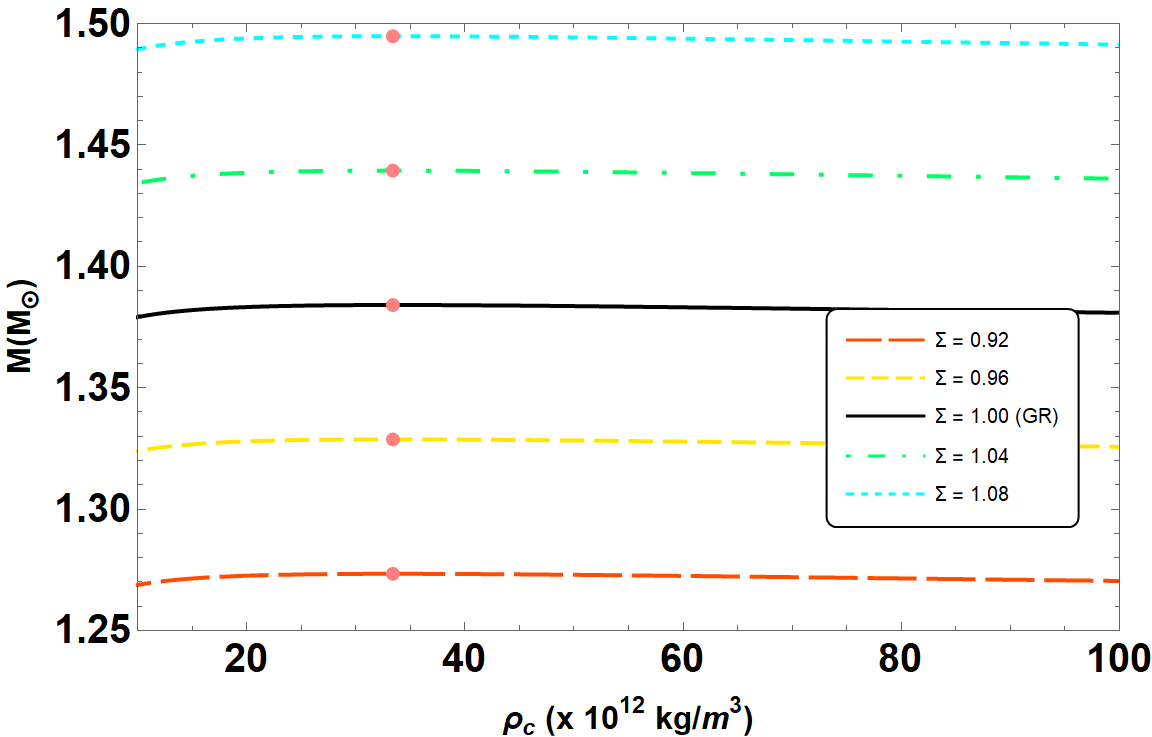}
   \includegraphics[width = 8 cm]{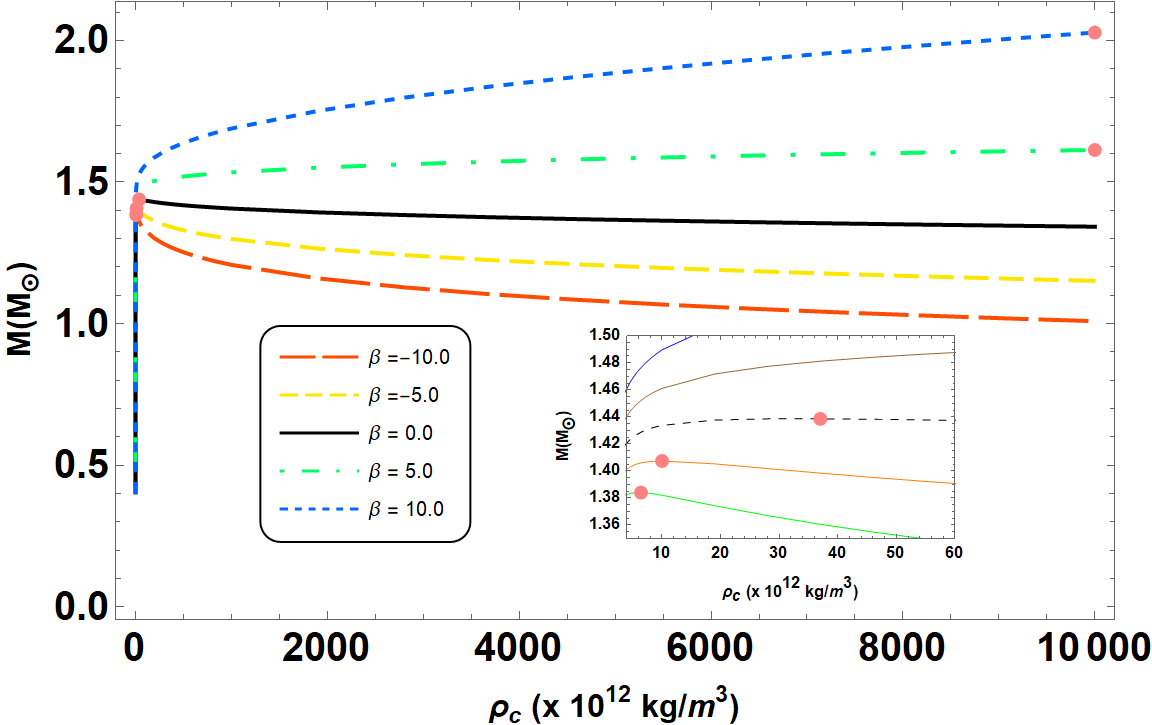}
    \caption{Star mass versus the central energy density $\rho_c $ for various values of $\Sigma$ and $\beta$. In figures, the pink points represent a boundary wall that separates the stable configuration region indicated by $dM/d\rho_c >0$ from the unstable one.}
    \label{stability1}
\end{figure}

At this point, we have studied the \textit{static stability criterion} \cite{harrison,ZN} to examine the 
 behavior of the equilibrium configuration under consideration. However, this is a necessary condition but not sufficient for confirming the stability of a spherical body.  Mathematically, this criterion has been extensively utilized in modified gravity theories also, see Refs. \cite{Maulana:2019sgd,Pretel:2022plg,Gammon:2023uss} and therein.  We express the definition through the following inequalities:
\begin{eqnarray}
\frac{d M}{d \rho_c} < 0 &~ \rightarrow \text{indicating an unstable configuration}, \\
\frac{d M}{d \rho_c} > 0 &~ \rightarrow \text{indicating a stable configuration}.
\label{criterion_M_rho_c}
\end{eqnarray}
The $M-\rho_c$ relations are shown in Fig. \ref{stability1}
with the parameter set provided in Figs. \ref{vary_Sigma} and  \ref{vary_beta}, respectively. As shown in Fig. \ref{stability1},
the mass of the stars increases with the central density, until it reaches a maximum value, and after that it starts decreasing. Although this second part of the curve is not physical, as it corresponds to instability according to the Harrison-Zeldovich criterion i.e., the stable configuration exists in the region where $dM/d\rho_c >0$. Consequently, the turning point indicated by the pink circles defined by $dM/d\rho_c =0$. 


\subsection{Adiabatic Indices}


\begin{figure}
    \centering
    \includegraphics[width = 8 cm]{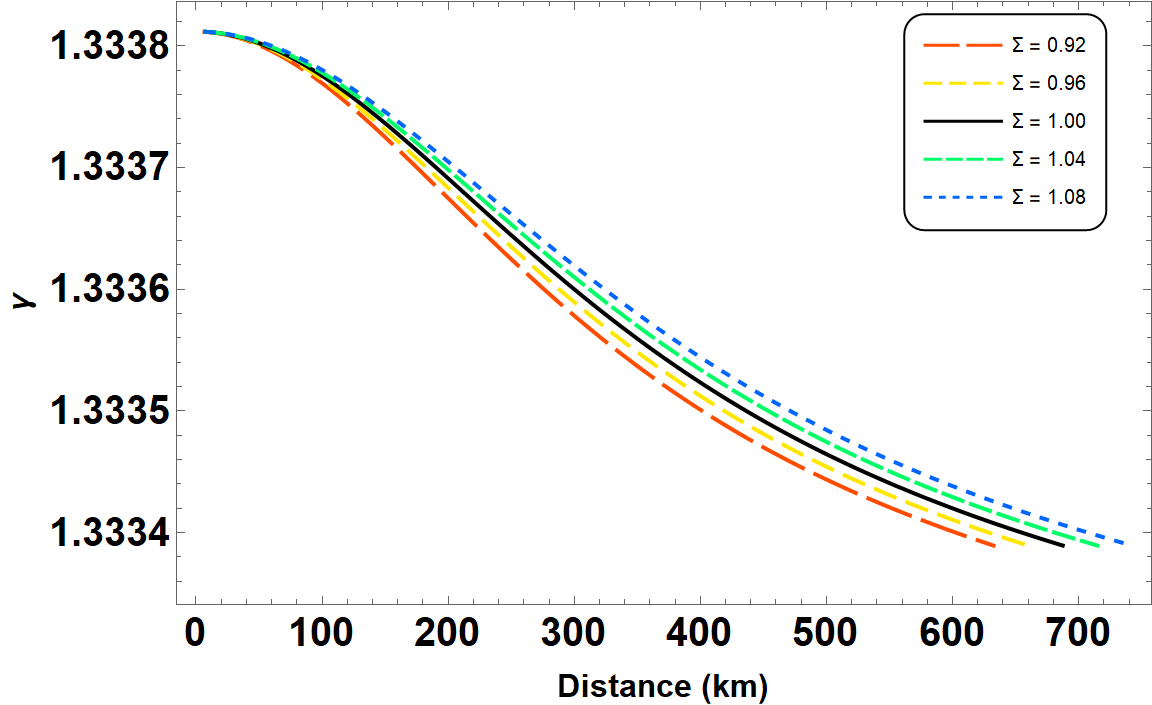}
    \includegraphics[width = 8 cm]{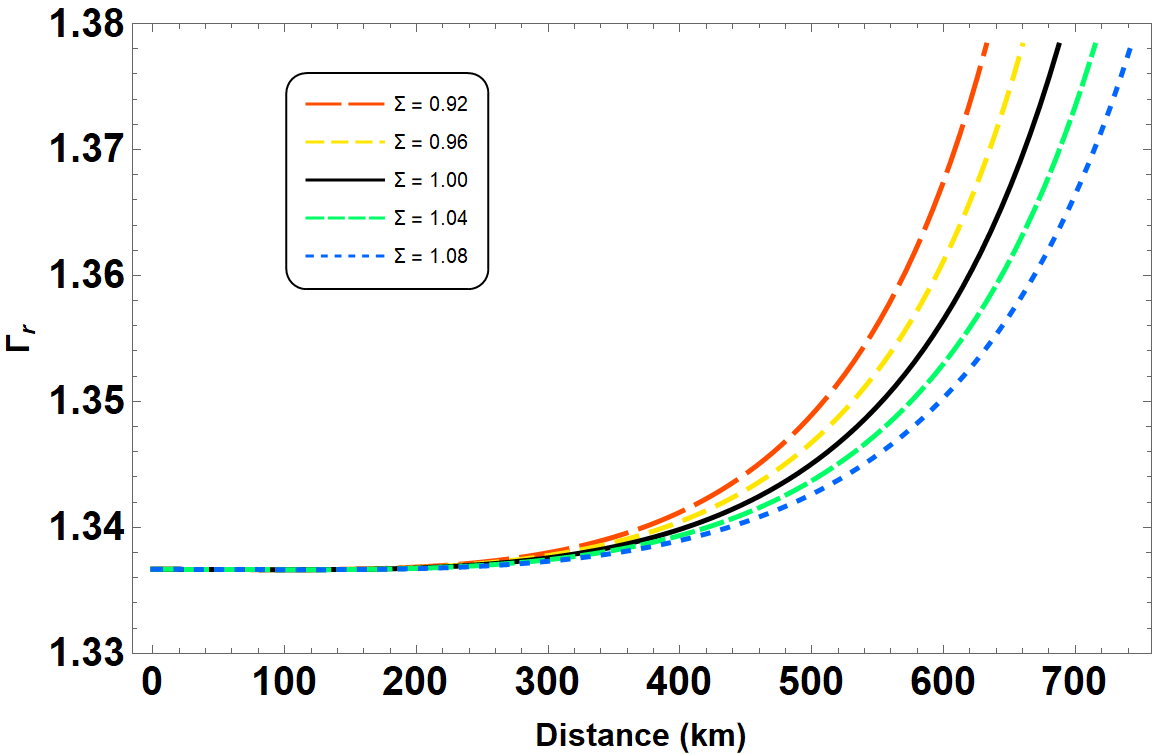}
    \includegraphics[width = 8 cm]{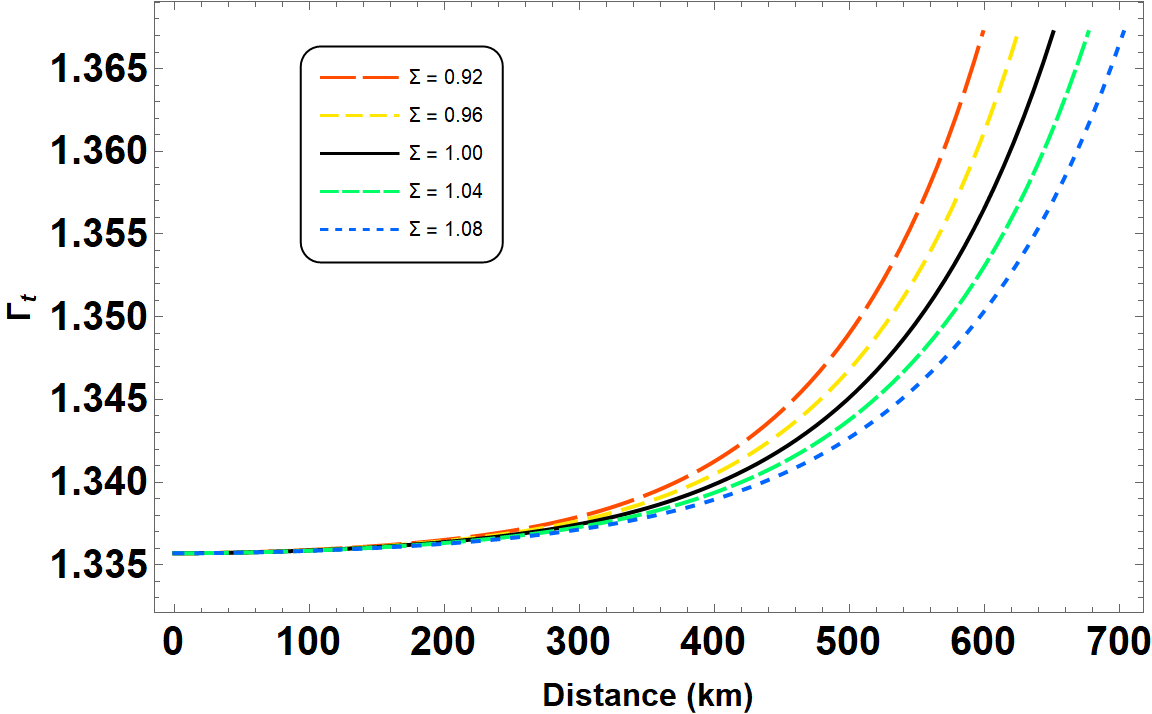}
    \caption{The adiabatic indices in the radial and tangential directions with a central density of $\rho_c = 3.7 \times 10^{13}$ kg/m$^3$ or 0.0208 MeV/fm$^3$, along with the parameter set depicted in Fig. \ref{vary_Sigma}. }
    \label{fig_adia_vary_sigma}
\end{figure}


\begin{figure}
    \centering
    \includegraphics[width = 8 cm]{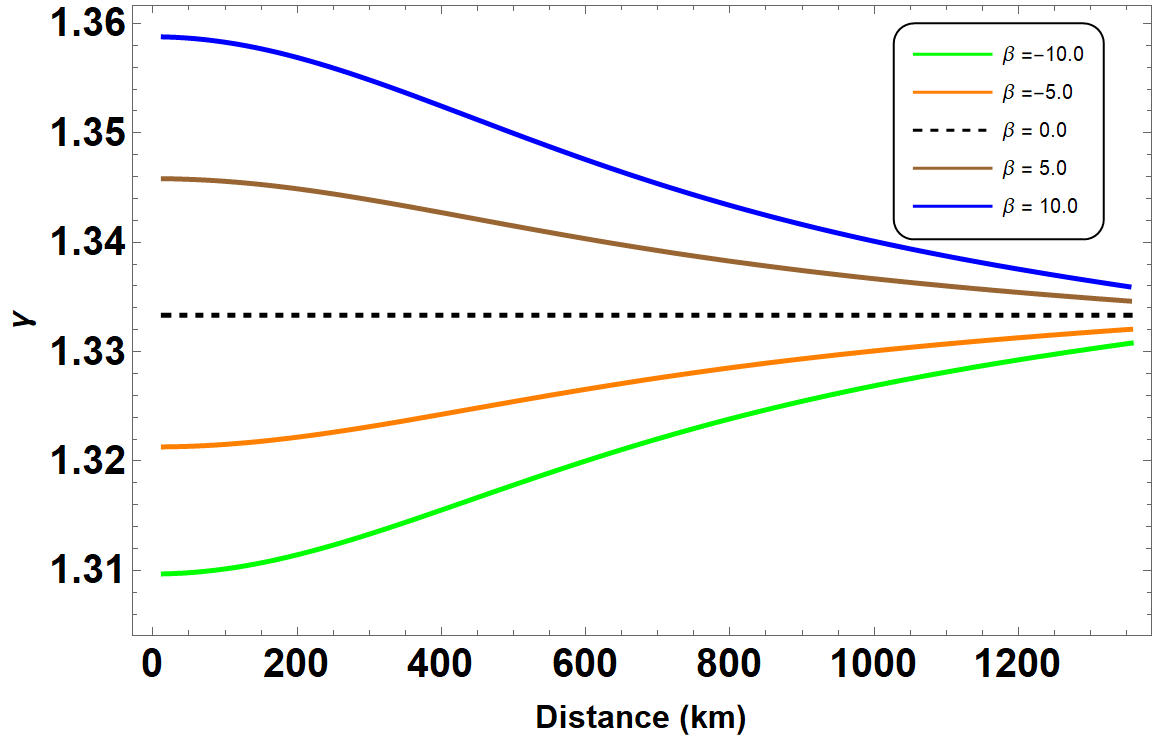}
    \includegraphics[width = 8 cm]{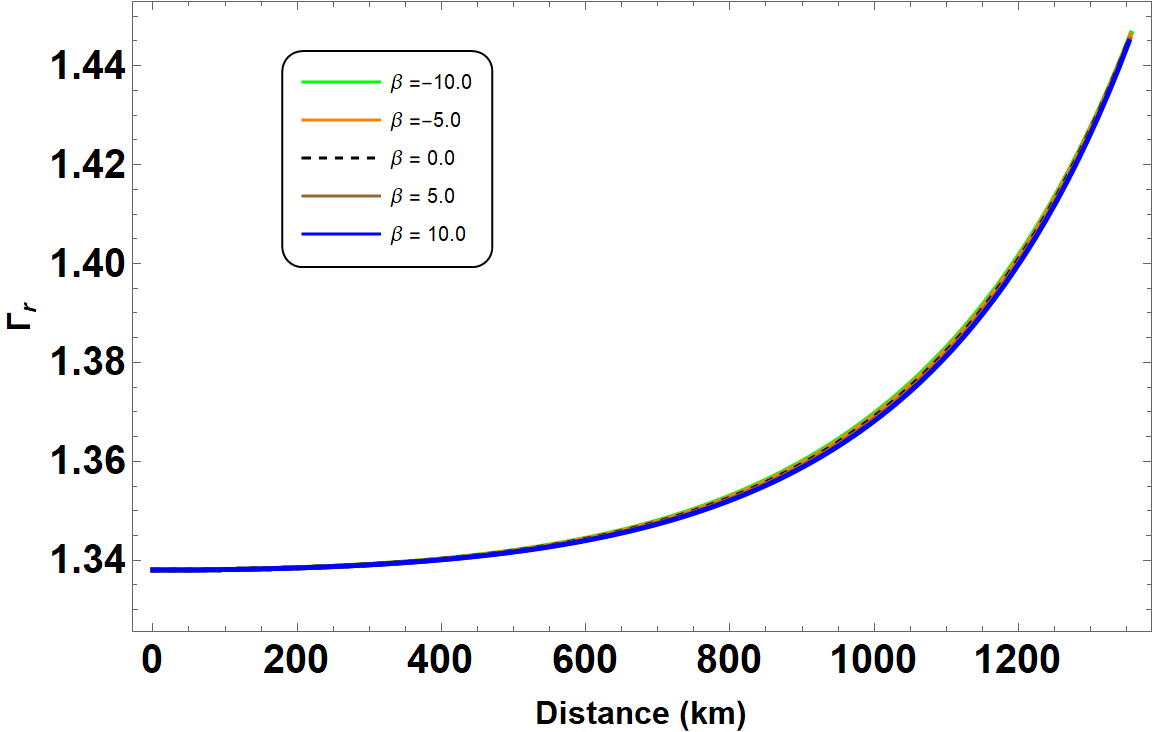}
    \includegraphics[width = 8 cm]{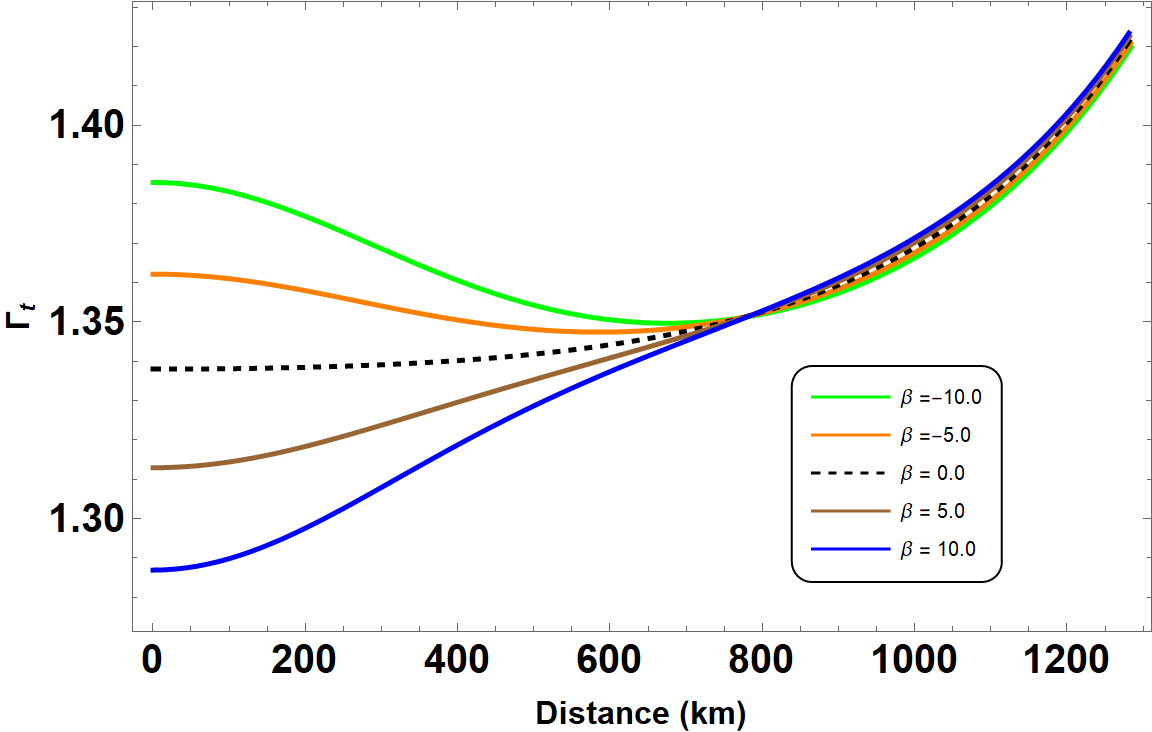}
    \caption{The adiabatic indices in the radial and tangential directions with a central density of $\rho_c = 5 \times 10^{12}$ kg/m$^3$ or 0.002814 MeV/fm$^3$, along with the parameter set depicted in \ref{vary_beta}.}
    \label{fig_adia_vary_beta_at_144}
\end{figure}


To investigate the dynamical stability of the configuration, in this section, we are going to study the adiabatic index ($\gamma$). It is important to note that Chandrasekhar \citep{Chandrasekhar:1964zza}
established the foundation for understanding dynamical instability in compact stars. This work provides key insights into the limits of the stability of relativistic stars. Mathematically,  the speed of sound, $c_s^2$ and the relativistic adiabatic index, $\Gamma$, of a fluid are defined by
\citep{Chandrasekhar:1964zza,Herrera:2012uyt}
\begin{equation}
c_s^2 \equiv \frac{dp}{d \rho}, \; \; \; \; \;  \Gamma \equiv c_s^2 \left( 1+\frac{\rho}{p}\right)
\end{equation}
while if the fluid is anisotropic, there is a sound speed and a relativistic adiabatic index for each direction, namely radial and tangential. The stability criterion based on $\Gamma$ is the following: its mean value must be larger than a critical value
\begin{equation}
\langle \Gamma \rangle \geq \Gamma_{cr}
\end{equation}
where the critical value is given by \cite{Moustakidis:2016ndw}
\begin{equation}
\Gamma_{cr} = \frac{4}{3} + \frac{19 M}{21 R}
\end{equation}
while the mean value is computed by \cite{Moustakidis:2016ndw}
\begin{equation}
\langle \Gamma \rangle = \frac{\int_0^R dr \: \Gamma(r) p(r) r^2 e^{\lambda+3 \Phi}}{\int_0^R dr \: p(r) r^2 e^{\lambda+3 \Phi}} .
\end{equation}
 The interested reader may consult for instance \cite{Moustakidis:2016ndw} for more details on the role of the relativistic adiabatic index on stability of stars.

\begin{eqnarray}\label{eq:adiabatic}
{\gamma}&=&\frac{4}{3}\left(1+\frac{{ \sigma}}{r|{  p}'_r|}\right)_{max}, \qquad
{\Gamma_r}=\frac{{\rho }+{p_r}}{{p_r}}{v_r^2}, \nonumber \\
 {\Gamma_t}&=&\frac{{\rho }+{p_t}}{{p_t}}{v_t^2 ,}
\end{eqnarray}
where $\Gamma_r$ and $\Gamma_t$ are the adiabatic indices in the radial and tangential direction, respectively. Since, $\Delta=0$ represents the isotropic solution, we have $\gamma=4/3$. In the case of soft-anisotropy where $\sigma <0 $,  we find $\gamma <4/3$, similar to what is observed in Newtonian theory. While for strong anisotropic case i.e., when $\sigma >0 $, we have $\gamma >4/3$. Neutral equilibrium occurs for $\Gamma=\gamma$ and stable equilibrium requires $\Gamma> \gamma$, see Refs \cite{Herrera:2012uyt,Heintzmann1975} for more details. The anisotropic star model exhibits stale behavior when $\Gamma_r > \gamma$ and $\Gamma_t > \gamma$ everywhere inside the star.  
 To discuss stability based on our calculation, we show in Fig. \ref{fig_adia_vary_sigma} and Fig. \ref{fig_adia_vary_beta_at_144} the variation of adiabatic indices $\Gamma_r$ and $\Gamma_t$ in both directions, namely radial and tangential, for $\Sigma$ variations and $\beta$ variations, respectively, assuming a certain value of the central energy density. Regarding $\Sigma$ variations, both indices increase with the radial coordinate, while at the same time they remain higher than the Newtonian value $4/3$. Regarding $\beta$ variations close to the Chandrasekhar limit, $M \sim 1.4 M_{\odot}$, the radial adiabatic index increases with the radial coordinate remaining higher than $4/3$, whereas in the tangential direction it exhibits a behavior that depends on the sign of the parameter $\beta$. In particular, a positive anisotropic factor implies instability, as $\Gamma_t < 4/3$.

\subsection{Sound Speed and Causality}


\begin{figure}
    \centering
    \includegraphics[width = 8 cm]{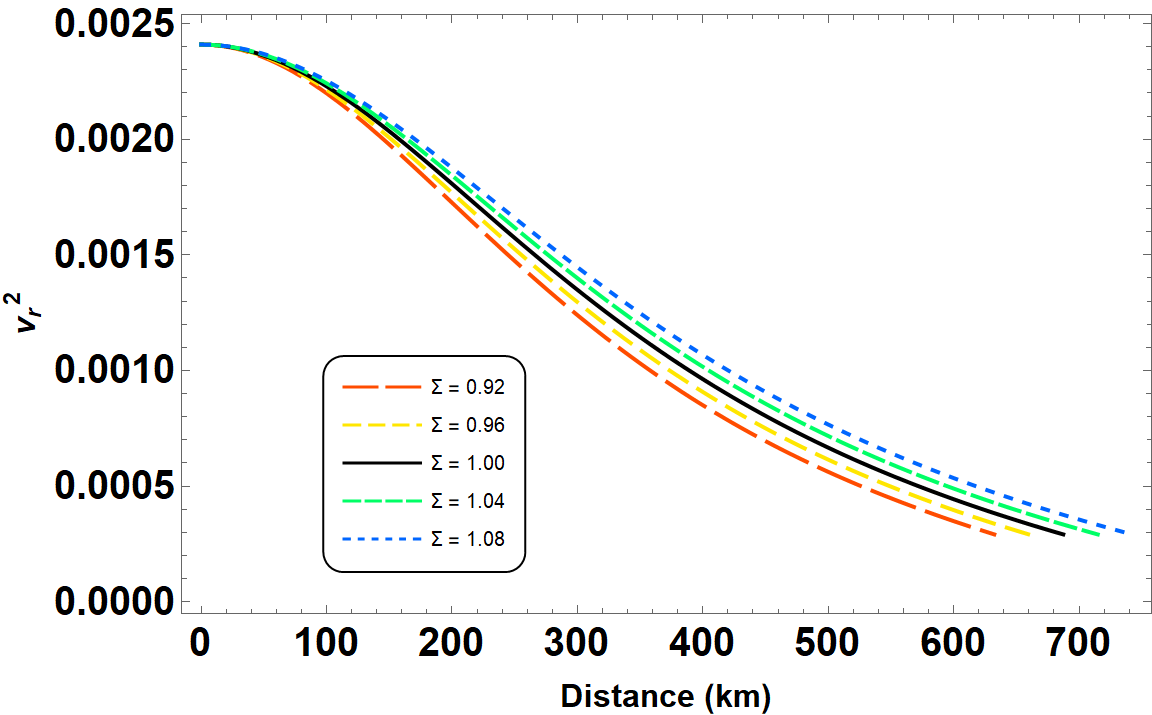}
    \includegraphics[width = 8 cm]{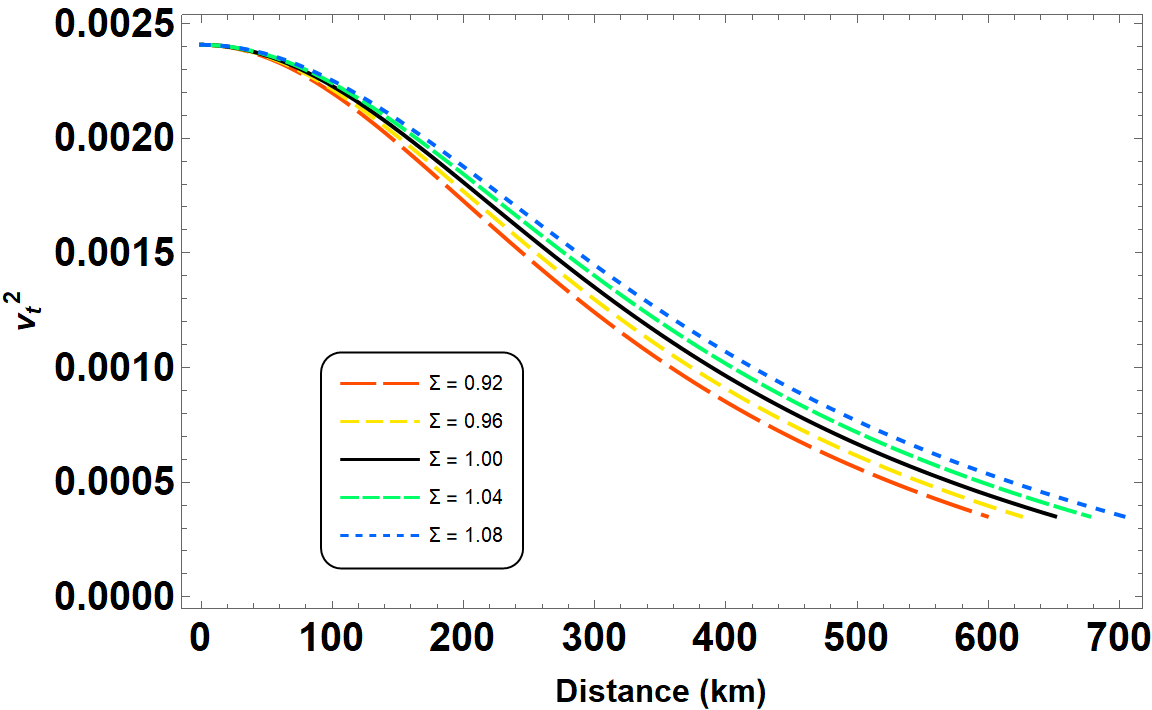}
    \caption{The squares of the sound speeds in the radial and tangential directions as a function of the distance from the center of the star to its edge, with parameters specified in Fig. \ref{vary_Sigma}.}
    \label{fig_vel_vary_sigma}
\end{figure}


\begin{figure}
    \centering
    \includegraphics[width = 8 cm]{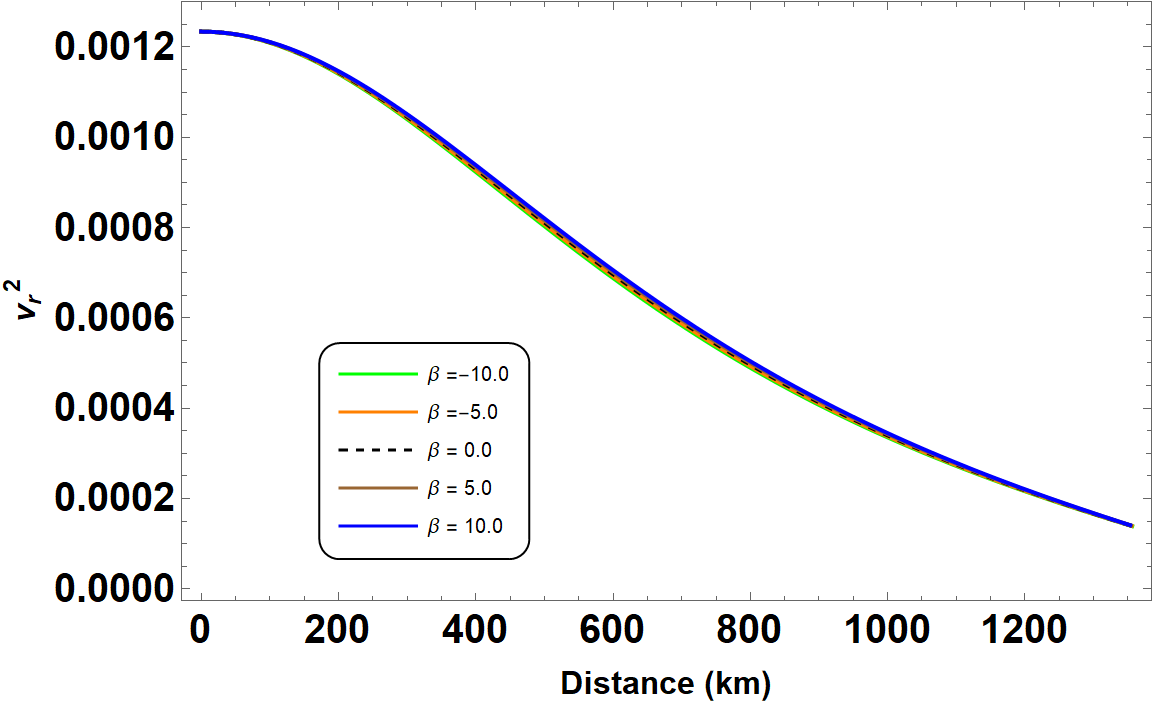}
    \includegraphics[width = 8 cm]{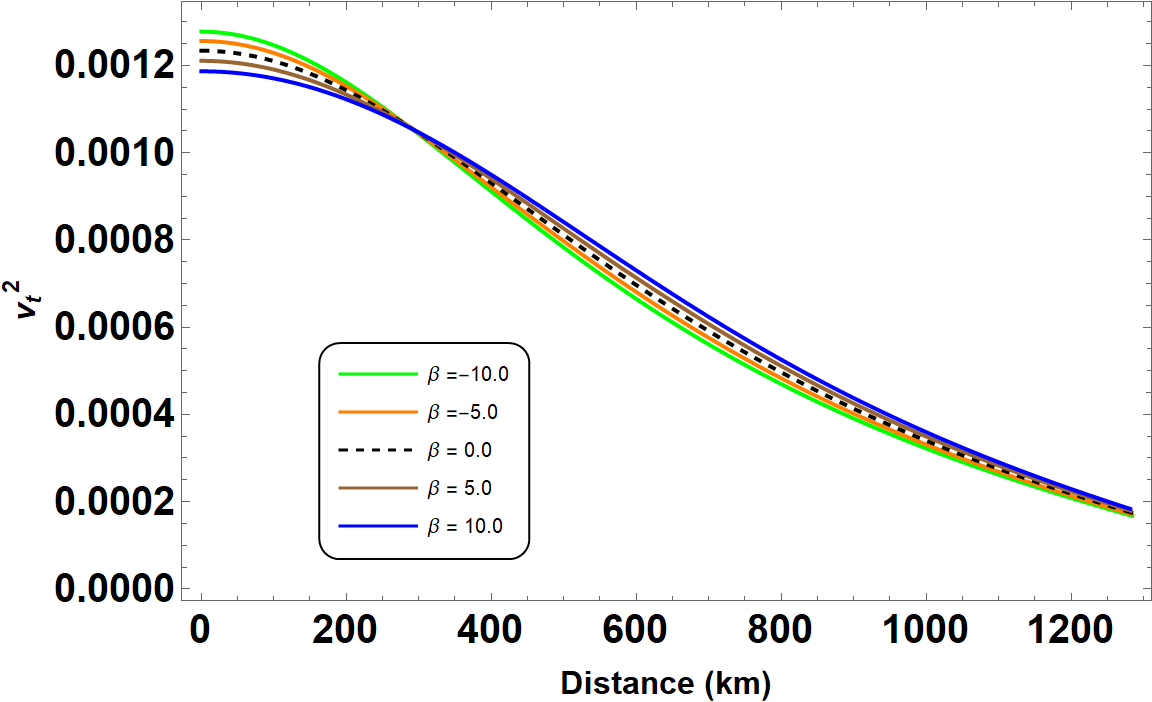}
    \caption{The squares of the sound speeds in the radial and tangential directions as a function of the distance from the center of the star to its edge, with parameters specified in  Fig. \ref{vary_beta}.}
    \label{fig_vel_vary_beta}
\end{figure}


We now turn to consider an additional validation, the squared speed of sound, defined as $v_{r,t}^2 = \frac{dp_{\{r,t\}}}{d\rho}$, serves as an essential validation to evaluate the physical correctness of our model. For Chandrasekhar's EoS, we calculate the sound speed both radially and tangentially within the stellar interior.  Finally, in Fig. \ref{fig_vel_vary_sigma} and Fig. \ref{fig_vel_vary_beta} we display the speed of sound in both directions, radial and tangential, for $\Sigma$ variations and $\beta$ variations, respectively, assuming a certain value of the central energy density. We observe that in all cases the sound speed monotonically decreases with the radial coordinate throughout the stars, remaining always lower than 1, and therefore causality is never violated.

\section{Conclusion}\label{sec6}

To summarize our work, in the present article we have studied in detail the properties of anisotropic White Dwarf stars within Rainbow gravity. First we presented the modified structure equations, and after that we briefly discussed the EoS for an ideal Fermi gas at zero temperature, which corresponds to the Chandrasekhar model. The model discussed here is characterized by two free parameters, namely $\beta$ and $\Sigma$. The first one is related to the anisotropic factor, while the other one measures deviations from GR. We summarized our main numerical results in a number of two Tables and seven figures, observing the impact of the two free parameters on the properties of the WD stars. Our findings indicate that a positive $\beta$ makes stars more massive, a negative $\beta$ makes WDs less massive, a $\Sigma < 1$ implies less massive objects, and a $\Sigma > 1$ implies more massive WD stars. Furthermore, causality is always respected as both speed of sounds decrease with the radial coordinate, and they always remain bounded $0 < c_s < 1$ throughout the WDs. Moreover, regarding the relativistic adiabatic index, the radial one in all cases increases with the radial coordinate, while at the same time it always remains higher than the Newtonian value $4/3$. The most striking observation reported in the present work for the first time is that anisotropic WDs characterized by a positive anisotropic factor close to the Chandrasekhar limit seem to become unstable, according to the criterion based on the relativistic adiabatic index along the tangential direction.

\section*{date availability}

There are no new data associated with this article.

\begin{acknowledgments}
Takol Tangphati was supported by Walailak University under the New Researcher Development scheme (Contract Number WU67268).

\end{acknowledgments}\

\end{document}